\documentclass[aip,jcp,reprint,amsmath,amssymb]{revtex4-2}
\usepackage{graphicx}
\usepackage{natmove}
\newcommand{\vv}[1]{\mathbf{#1}}
\renewcommand{\d}[1]{\ensuremath{\operatorname{d}\!{#1}}}

\begin{document}
\title{Transport properties of monodisperse and bidisperse hard-sphere colloidal suspensions from multiparticle collision dynamics simulations}

\author{Michael P. Howard}
\email{mphoward@auburn.edu}
\affiliation{Department of Chemical Engineering, Auburn University, Auburn, AL 36849, USA}

\begin{abstract}
The shear viscosities, long-time self-diffusion coefficients, and sedimentation velocities in monodisperse and bidisperse hard-sphere colloidal suspensions are simulated for volume fractions up to 0.40 using multiparticle collision dynamics with a discrete particle model. The bidisperse suspensions have diameter ratios of 2 and 4 and equal amounts of each particle by volume. All measured properties for monodisperse suspensions are found to be in good agreement with prior literature; however, they highlight the sensitivity of the simulation method to discretization effects. The sedimentation velocities for the bidisperse suspensions are also in reasonable agreement with prior literature, including direction reversal for the smaller particles when the diameter ratio is 4. This work provides reference data for transport properties of colloidal suspensions and establishes the suitability of multiparticle collision dynamics for modeling suspensions of particles with different sizes.
\end{abstract}

\maketitle

\section{Introduction}
Polydispersity is a practical reality of many colloidal suspensions that nontrivially modifies their rheology \cite{luckham:jcis:1999, rastogi:jcp:1996, pednekar:jor:2018} and has important implications for industrial processes such as dip coating \cite{jeong:jfm:2022}. Polydispersity can also be exploited to engineer mesoscale structures, and their resulting properties, in materials \cite{fernando:iecr:2025}. For example, hard-sphere colloidal particles will spontaneously segregate by size during fast solvent drying \cite{fortini:prl:2016, howard:lng:2017}, and particle interactions \cite{zhou:prl:2017} and dynamics \cite{sear:pre:2017, statt:jcp:2018, howard:jcp:2020, liu:acsnano:2019, park:sm:2022} play a critical in determining the structure that forms. This effect has been used to produce coatings with functional properties such as abrasion-resistance \cite{tinkler:jcis:2021, kargarfard:poc:2021}, bacteria-resistance \cite{dong:acs:2020}, or controlled release \cite{le:iecr:2026} as well supraparticles with engineered porosity \cite{liu:acsnano:2019b}.

Despite being the simplest case of polydispersity, the transport properties of bidisperse colloidal suspensions are far less well-characterized than those of monodisperse colloidal suspensions. Experimental measurements of the shear viscosity \cite{rodriguez:lng:1992} and particle sedimentation coefficients \cite{mirza:ces:1979, alnaafa:cjce:1989, alnaafa:aichej:1992, davis:aichej:1988} have been reported for bidisperse suspensions, but the data is sparse and so difficult to interpolate for specific particle sizes and concentrations of interest. Particle-based simulations that faithfully capture solvent-mediated hydrodynamic interactions in the suspension are potentially promising for filling these gaps \cite{bolintineanu:cpm:2014, howard:coce:2019}. However, bidisperse suspensions are more challenging to simulate than monodisperse suspensions because disparity in particle size typically increases the number of particles that must be modeled and the computational costs of the simulations \cite{howard:cpc:2016}. Stokesian dynamics simulations \cite{brady:jfm:1988} have been used to characterize short-time transport properties for bidisperse suspensions \cite{wang:jcp:2015, wang:jcp:2015b}, but they are computationally demanding \cite{wang:jcomputphys:2016, fiore:jfm:2019} and so can be difficult to use routinely to study long-time or large-scale behaviors.

The purpose of this article is to investigate the suitability of multiparticle collision dynamics \cite{malevanets:jcp:1999, gompper:2009, howard:coce:2019} (MPCD) as a method for simulating bidisperse colloidal suspensions. MPCD uses an explicit, highly simplified solvent that is straightforwardly coupled to colloidal particles \cite{poblete:pre:2014}, and massively parallel open-source software is available to perform MPCD simulations efficiently \cite{howard:cpc:2018, westphal:cpc:2025}. I have recently shown that MPCD produces reasonable results for the transport properties of not only monodisperse spherical particles \cite{wani:jcp:2022} but also shape-anisotropic particles such as cubes, octahedra, tetrahedra, and spherocylinders \cite{wani:sm:2024}. MPCD has also been used to simulate mixtures of spherical and rod-like particles \cite{yetkin:lng:2023}, but to my knowledge, it has not yet been applied to bidisperse suspensions of spherical particles. It is important to establish and assess the transport properties of bulk suspensions in MPCD before applying the method to more complex dynamic processes such as self-assembly.

In this article, I use MPCD to simulate the shear viscosity, long-time self-diffusion coefficients, and sedimentation velocities in monodisperse and bidisperse hard-sphere colloidal suspensions as a function of volume fraction. I use particles with three different sizes, giving diameter ratios of 2 and 4 for the bidisperse suspensions. The measured properties are generally in good agreement with theoretical expectations and previous simulations where available, demonstrating the suitability of MPCD for simulating suspensions of particles with different sizes. However, they also highlight the sensitivity of the MPCD method to discretization effects that should be carefully considered.

\section{Model and Methods}
\label{sec:methods}
In the following, all quantities will be reported in a consistent system of units where $m$ is the unit of mass, $\ell$ is the unit of length, and $\varepsilon$ is the unit of energy. The unit of time is $\tau = \sqrt{m\ell^2/\varepsilon}$ and the unit of temperature is $\varepsilon/k_{\rm B}$, where $k_{\rm B}$ is the Boltzmann constant. All simulations were performed using HOOMD-blue \cite{anderson:cms:2020, howard:cpc:2018, howard:cpc:2016, howard:cms:2019} extended with azplugins \cite{azplugins}. The versions used for each calculation are noted.

\subsection{Model}
The solvent was modeled using MPCD. The solvent particles had mass $1\,m$, and their number density was $5/\ell^3$. In MPCD, solvent particles do not have pairwise interactions with each other; instead, they evolve in alternating streaming and collision steps. The solvent particles moved according to Newton's equations of motion for $0.1\,\tau$, then exchanged momentum with other particles in cubic cells with edge length $1\,\ell$ according to the stochastic rotation dynamics rule without angular-momentum conservation \cite{malevanets:jcp:1999}. The collision angle was $130^\circ$, and a rotation axis was randomly chosen for each cell from the unit sphere. A cell-level Maxwell--Boltzmann thermostat \cite{huang:jcp:2010, huang:pre:2015} was used to maintain constant temperature $1\,\varepsilon/k_{\rm B}$, and the grid of collision cells was shifted before each collision along each Cartesian axis by a uniform random amount drawn in the interval between $\pm 0.5\,\ell$ to ensure Galilean invariance \cite{ilhe:pre:2001, ilhe:pre:2003}.

\begin{figure}[b]
    \centering
    \includegraphics{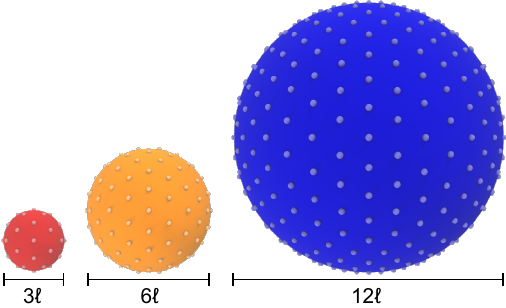}
    \caption{Particles with diameter $3\,\ell$, $6\,\ell$, and $12\,\ell$ along with their discretized surfaces. The number of surface particles for each was 42, 162, and 642, corresponding to surface densities of $1.49\,\ell^{-2}$, $1.43\,\ell^{-2}$, and $1.42\,\ell^{-2}$, respectively. The images were rendered using VMD 1.9.3 \cite{humphrey:jmg:1996}.}
    \label{fig:particles}
\end{figure}

The colloidal particles were modeled as nearly hard spheres using the core-shifted Weeks--Chandler--Andersen pairwise interaction potential \cite{weeks:jcp:1971},
\begin{widetext}
\begin{equation}
u(r) = \begin{cases}
\displaystyle 4\varepsilon\left[\left(\frac{\sigma}{r-\Delta_{ij}}\right)^{12} - \left(\frac{\sigma}{r-\Delta_{ij}}\right)^6 + \frac{1}{4}\right],& r \le \Delta_{ij} + 2^{1/6}\sigma \\
0,& {\rm otherwise}
\end{cases},
\label{eq:wca}
\end{equation}
\end{widetext}
where $r$ is the distance between the centers of two particles, $\sigma = 1\,\ell$ sets the length scale for the repulsion, and $\Delta_{ij} = (d_i + d_j)/2 - \sigma$ shifts the divergence of the repulsion based on the diameters of the particles, denoted $d_i$ for particle type $i$. Three types of particles having diameter $3\,\ell$, $6\,\ell$, and $12\,\ell$ were considered (Fig.~\ref{fig:particles}).

The colloidal particles were coupled to the solvent through the collision step using a discrete particle model \cite{poblete:pre:2014}. The surfaces of the colloidal particles were discretized by iteratively subdividing the faces of an icosahedron and rescaling its vertices to lie on the sphere to achieve roughly the same surface density (Fig.~\ref{fig:particles}). These surface sites were connected to their nearest neighbors on the surface and to a central site, used to apply Eq.~\eqref{eq:wca}, by harmonic springs,
\begin{equation}
u_{\rm b}(r) = \frac{k}{2}(r-r_0)^2,
\end{equation}
with a stiff spring constant $k = 5000\,\varepsilon/\ell^2$ and preferred length $r_0$ rounded to three decimal places. The surface sites were assigned mass $5\,m$ and participated in the collisions with the solvent \cite{malevanets:epl:2000}. The central site was also assigned a mass of $5\,m$ but did not interact directly with the solvent. Between collisions, all sites comprising a colloidal particle moved according to Newton's equations of motion using conventional molecular dynamics (MD) methods with a Verlet integration scheme (timestep $0.002\,\tau$). This approach has been previously shown to reasonably capture the dynamics of single spherical particles \cite{poblete:pre:2014, peng:jcp:2024}, as well as the long-time self-diffusion and sedimentation coefficients of monodisperse spherical \cite{wani:jcp:2022} and shape-anisotropic \cite{wani:sm:2024} particles.

I considered monodisperse suspensions of all three particle types ($d_1 = 3\ell$, $6\,\ell$, or $12\,\ell$) and bidisperse suspensions with $d_1 = 3\,\ell$ and $d_2 = 6\,\ell$ or $12\,\ell$. The simulation box was a cube with edge length $L = 120\,\ell$ and periodic boundary conditions. A certain number of particles $N_i$ of each type $i$ were dispersed into the box to achieve a desired volume fraction $\phi_i = N_i \pi d_i^3/(6L^3)$ of that type. The particles were placed by randomly selecting sites from a face-centered cubic lattice commensurate with the box and having a lattice constant of roughly $\sqrt{2}\,d_i$. For the bidisperse suspensions, the larger particles were placed first, then the lattice sites for the smaller particles giving hard-sphere overlap with the larger particles were removed. The total volume fraction $\phi = \sum_i \phi_i$ was varied between 0.01 and 0.40. I considered only equal-volume ($\phi_1 = \phi_2$) bidisperse suspensions because prior Stokesian dynamics simulations \cite{wang:jcp:2015} showed more sensitivity of transport properties to the diameter ratio $d_2/d_1$ and the total volume fraction $\phi$ than to the relative composition, $\phi_1/\phi$.

\subsection{Equilibrium structure}
The initial configurations were equilibrated for $10^3\,\tau$ using isothermal--isochoric MD with a Bussi thermostat \cite{bussi:jcp:2007} (time constant $0.1\,\tau$) to maintain constant temperature $1\,\varepsilon/k_{\rm B}$. Hydrodynamic interactions do not affect equilibrium structure, so only the central sites were simulated using a larger timestep of $0.005\,\tau$. After equilibration, particle configurations were recorded every $10\,\tau$ until a certain number were collected ($10^3$, $10^4$, and $10^5$ for the monodisperse suspensions with $d_1 = 3\,\ell$, $6\,\ell$, and $12\,\ell$; $10^4$ and $10^5$ for the bidisperse suspensions with $d_2/d_1 = 2$ and 4). Five of these configurations, taken at equal time intervals, served as initial configurations for subsequent simulations with hydrodynamic interactions. These simulations were performed using HOOMD-blue version 4.8.2.

The partial static structure factor for particles of types $i$ and $j$ was computed from all recorded configurations using the Ashcroft--Lengreth normalization convention \cite{ashcroft:pr:1967},
\begin{equation}
S_{ij}(\vv{q}) = \frac{1}{\sqrt{N_i N_j}} \left\langle\sum_{n=1}^{N_i} \sum_{p=1}^{N_j} e^{-i \vv{q} \cdot (\vv{r}_n^{(i)} - \vv{r}_p^{(j)})}\right\rangle,
\end{equation}
where $\vv{r}_n^{(i)}$ is the position of a given particle $n$ of type $i$, $\vv{q}$ is a wavevector commensurate with the periodic boundaries of the simulation box, and the angle brackets denote an ensemble average. Under this convention, the total static structure factor is
\begin{equation}
S(\vv{q}) = \sum_{i,j} \sqrt{x_i x_j} S_{ij}(\vv{q}),
\end{equation}
where the double sum runs over all particle types $i$ and $j$, and $x_i = N_i / N$ is the number fraction of type $i$ with $N = \sum_i N_i$ being the total number of particles. The magnitude of the smallest wavevector commensurate with the simulation box was $\Delta q = 2\pi/L$. I calculated the isotropically averaged $S_{ij}(q)$ for wavevector magnitudes $\Delta q \le q \le 20 \Delta q$ using freud \cite{ramasubramani:cpc:2020} (version 3.1.0) with a bin spacing of $\Delta q$ for wavevector averaging (Figs.~S1 and S5).

\subsection{Shear viscosity}
I next measured the shear viscosity $\eta$ of the suspensions using reverse nonequilibrium simulations \cite{muller:physrev:1999, tenney:jcp:2010}. The surfaces of the equilibrated particles were discretized, and solvent was added randomly everywhere in the simulation box with velocities randomly thermalized at temperature $1\,\varepsilon/k_{\rm B}$. For only these simulations, the particle configuration was duplicated to make a simulation box that was twice as long in the $y$ direction (i.e., $2L = 240\,\ell$). Shear flow was then generated using an artifical momentum-swapping procedure. Every $0.1\,\tau$, solvent particles were sorted into two planar slabs of thickness $1\,\ell$: a ``lower'' slab at the bottom of the simulation box and an ``upper'' slab in the middle of the simulation box in the $y$ direction (spaced apart by $L$). The particles having the $x$-component of their velocity closest to a target $0.5\,\ell/\tau$ in the lower slab and $-0.5\,\ell/\tau$ in the upper slab were identified, and the $x$-component of the velocity was exchanged between the first 100 pairs of particles sorted in this way. This procedure conserves both linear momentum and kinetic energy for the system, and with appropriate parameters, it produces two regions of opposing steady shear within the same simulation box. The selected parameters are expected to achieve a maximum flow speed of about $0.5\,\ell/\tau$ for a pure solvent giving a maximum shear rate less than about $0.01\,\tau^{-1}$. Thus, these simulations probe the low-shear regime, as verified in subsequent analysis discussed in Sec.~\ref{sec:results}.

The momentum exchange procedure was carried out for $2 \times 10^4\,\tau$ to allow the flow to reach steady state, then a production simulation of $2 \times 10^4\,\tau$ was performed to measure the shear viscosity. During the production simulation, the cumulative momentum exchanged by swapping $p_x(t)$ was recorded as a function of time $t$. The mass-averaged velocity of the suspension in the $x$ direction as a function of position in the $y$ direction, $u_x(y)$, was measured every $10\,\tau$ using a histogram of the solvent and all sites that comprised the particles in $y$ with bin width $0.5\,\ell$ (Figs.~S3 and S7). The average rate of momentum transfer $\dot p_x$ was computed over the entire production simulation. The shear rate $\dot\gamma = \d{v_x}/\d{y}$ was extracted by a linear fit of the measured velocity, excluding a region of thickness $24\,\ell$ centered around each momentum-exchange slab and applying the symmetry of the opposing shear flows. The shear viscosity $\eta$ was then computed as
\begin{equation}
    \eta = \frac{1}{2L^2} \frac{\dot p_{x}}{\dot\gamma}.
    \label{eq:viscosity-sim}
\end{equation}
I repeated this procedure for the first three equilibrated configurations, and I report the mean $\eta$ with uncertainty estimated as one standard error of the mean from these simulations. The simulations were performed using HOOMD-blue 5.3.1 with azplugins 1.1.0.

\subsection{Long-time self-diffusion coefficient}
The long-time self-diffusion coefficient $D_i$ for particles of type $i$ was determined from equilibrium simulations. Discretized, solvated initial particle configurations were prepared and equilibrated for $10^3\,\tau$, then a production simulation of $5 \times 10^4\,\tau$ was performed during which the particle positions were recorded every $10\,\tau$. The average mean squared displacement $\langle\Delta r_i^2(t)\rangle$ of a particle of type $i$ after time $t$ was computed up to $2 \times 10^4\,\tau$ using all configurations as independent time origins and averaging over all particles. The long-time self-diffusion coefficient was then computed as
\begin{equation}
D_i = \frac{1}{6} \lim_{t\to\infty} \frac{\d{\langle \Delta r_i^2\rangle}}{\d{t}}.
\end{equation}
The time derivative of $\langle\Delta r_i^2\rangle$ was evaluated numerically (Figs.~S4 and S8), and the long-time limit was taken by averaging its value for $t \ge 10^4\,\tau$.

Self-diffusion coefficients simulated in cubic boxes with periodic boundary conditions are known to be reduced compared to their values in an infinitely large box. This finite-size effect can be corrected by adding \cite{dunweg:jcp:1993, yeh:jpc:2004}
\begin{equation}
\Delta D = \xi \frac{k_{\rm B} T}{6\pi\eta L}
\label{eq:D-finite}
\end{equation}
to the simulated value, where $\xi = 2.837297$ is a constant related to summing long-ranged hydrodynamic interactions over a cubic lattice of simulation boxes \cite{hasimoto:jfm:1959}. The mean value and uncertainty (one standard error of the mean) of the self-diffusion coefficients were determined from five independent simulations, then corrected using Eq.~\eqref{eq:D-finite} with uncertainty propagation. The simulations were performed using HOOMD-blue 5.0.1.

\subsection{Sedimentation velocity}
\label{sec:methods:sediment}
The sedimentation velocity $U_i$ for particles of type $i$ was determined from force-driven nonequilibrium simulations. The same solvated initial configurations were used as for the self-diffusion coefficients, but now a body force $f_i$ was applied to particles of type $i$ in the positive $x$ direction to simulate sedimentation. I used a force of $0.5\,\varepsilon/\ell$ for the particles with diameter $6\,\ell$, and I made $f_i \sim d_i^3$ to be consistent with sedimentation resulting from a density mismatch. The body force was distributed equally between the central and surface sites. A counterforce was also applied to all solvent particles to ensure the suspension was force free. Sedimentation was simulated for $10^3\,\tau$ to reach steady state. The sedimentation velocity $U_i$ averaged over all particles of type $i$ was then recorded every $0.102\,\tau$ during a $5 \times 10^4\,\tau$ production simulation and averaged over all measurements. The reason for this sampling rate is to collect samples at different offsets relative to the collision period.

Similarly to diffusion coefficients, sedimentation velocities simulated in cubic boxes with periodic boundary conditions are also reduced compared to their values in an infinitely large box. For monodisperse suspensions, the correction added to the simulated value is \cite{mo:pf:1994}
\begin{equation}
\Delta U_1 = \frac{\xi}{6 \pi \eta L} S(0) f_1,
\label{eq:U-finite-mono}
\end{equation}
where $S(0) = \lim_{q\to 0} S(q)$ is the limit of the static structure factor approaching the zero wavevector. This value was extracted from the equilibrium simulation data by fitting to a quadratic form $c_1 + c_2 q^2$ for $0.15\,\ell^{-1} \le q \le 0.5\,\ell^{-1}$ for diameters $3\,\ell$ and $6\,\ell$ and $q \le 0.3\,\ell^{-1}$ for diameter $12\,\ell$ (Fig.~S1) and extrapolating to $q=0$. For bidisperse suspensions, the corrections for each particle type are more complicated and involve the partial static structure factors \cite{wang:jcp:2015}
\begin{align}
\Delta U_1 &= \frac{\xi}{6 \pi \eta L} \left[S_{11}(0) f_1 + \sqrt{\frac{x_2}{x_1}} S_{12}(0) f_2 \right]  \\
\Delta U_2 &= \frac{\xi}{6 \pi \eta L} \left[\sqrt{\frac{x_1}{x_2}} S_{12}(0) f_1 + S_{22}(0) f_2\right],
\label{eq:U-finite-bi}
\end{align}
where $S_{ij}(0) = \lim_{q \to 0} S_{ij}(q)$. These values were fit and extrapolated in the same way as for the monodisperse suspensions using the fitting range for $q$ associated with the larger particle (Fig.~S5). As for the self-diffusion coefficients, the mean value and uncertainty (one standard error of the mean) of the sedimentation velocities were determined from five independent simulations, then corrected using Eqs.~\eqref{eq:U-finite-mono}--\eqref{eq:U-finite-bi} with uncertainty propagation. The simulations were performed using HOOMD-blue 5.0.1 with azplugins 1.1.0.

\section{Results and Discussion}
\label{sec:results}
\subsection{Monodisperse suspensions}
\label{sec:results:mono}
I started by characterizing the transport properties of the monodisperse suspensions. It is expected that transport properties for monodisperse suspensions of particles with different diameters should collapse as a function of the volume fraction $\phi$ when made nondimensional by their values in the dilute limit. I and others have previously simulated the self-diffusion and sedimentation of monodisperse hard-sphere colloidal suspensions using MPCD \cite{wani:jcp:2022, wani:sm:2024, peng:jcp:2024}, typically finding reasonable agreement with theory, experiments, and other simulation methods. However, it is also known that MPCD simulations can be sensitive to the size of the particles \cite{padding:pre:2006} as well as to their surface discretization \cite{poblete:pre:2014, peng:jcp:2024}, so it was important to establish whether theoretical consistency was obtained for the monodisperse suspensions before simulating their bidisperse mixtures.

\begin{figure}
    \centering
    \includegraphics{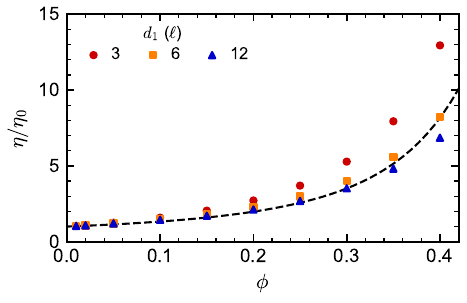}
    \caption{Simulated shear viscosity $\eta$ as a function of volume fraction $\phi$ for monodisperse suspensions of particles with diameter $d_1$. The dashed line is Eq.~\eqref{eq:viscosity-mono}.}
    \label{fig:viscosity-mono}
\end{figure}

I first considered the shear viscosity $\eta$ of the monodisperse suspensions (Fig.~\ref{fig:viscosity-mono}). The simulations should probe the low-shear limit, for which an approximation of the viscosity has been derived \cite{verberg:pre:1997},
\begin{equation}
\frac{\eta}{\eta_0} = g^+\left[1 + \frac{1.44 (\phi g^+)^2}{1-0.1241\phi+10.46\phi^2} \right].
\label{eq:viscosity-mono}
\end{equation}
In this approximation, $\eta_0$ is the viscosity of the pure solvent, and $g^+$ is the value of the hard-sphere radial distribution function at contact. I measured an MPCD solvent viscosity of $\eta_0 = 3.95\,\varepsilon\tau/\ell^3$ using the same reverse nonequilibrium simulation protocol as for the suspensions, which is in excellent agreement with the theoretical expectation \cite{kikuchi:jcp:2003, tuzel:physrev:2003, ripoll:physrev:2005} and previously reported measurements \cite{statt:prf:2019}, while $g^+$ was modeled using the Carnahan--Starling equation of state \cite{carnahan:jcp:1969},
\begin{equation}
g^+ = \frac{1-\phi/2}{(1-\phi)^3}.
\end{equation}
The simulated shear viscosity for the monodisperse suspensions was in generally good agreement with Eq.~\eqref{eq:viscosity-mono}. The viscosity for diameter $3\,\ell$ was consistently somewhat larger than expected, which is most obvious for the larger volume fractions. The viscosity for diameters $6\,\ell$ and $12\,\ell$ were in line with expectations, although the viscosity for diameter $12\,\ell$ was consistently smaller than for diameter $6\,\ell$. For self-consistency, the measured suspension viscosities were used for calculating finite-size corrections to the self-diffusion and sedimentation coefficients.

\begin{figure}
    \centering
    \includegraphics{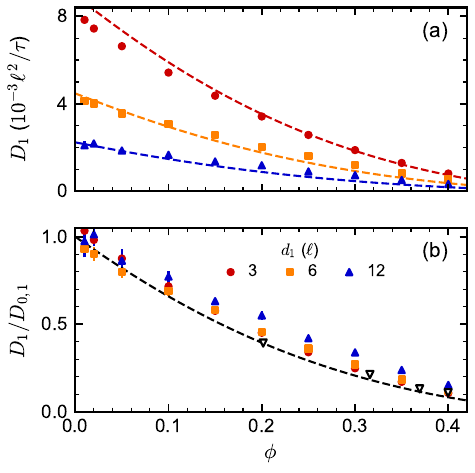}
    \caption{(a) Simulated long-time self-diffusion coefficient $D_1$ as a function of volume fraction $\phi$ for monodisperse suspensions of particles with diameter $d_1$. The dashed line is Eq.~\eqref{eq:diffusion-mono} with $D_{0,1}$ set to its theoretically expected value. (b) The same as (a) but the simulated $D_1$ is normalized by the simulated single-particle self-diffusion coefficient $D_{0,1}$. The open symbols are the Stokesian dynamics results of Ref.~\citenum{foss:jfm:1999}.}
    \label{fig:diffusion-mono}
\end{figure}

I next analyzed the long-time self-diffusion coefficient $D_1$ of the monodisperse suspensions (Fig.~\ref{fig:diffusion-mono}), which is theoretically expected to be \cite{tokuyama:pre:1994}
\begin{equation}
\frac{D_1}{D_{0,1}} = \frac{1 - 9\phi/32}{1 + H + (\phi/\phi_0)/(1-\phi/\phi_0)^2},
\label{eq:diffusion-mono}
\end{equation}
where $D_{0,1}$ is the long-time self-diffusion coefficient of a single particle,
\begin{equation}
H = \frac{2b^2}{1-b} - \frac{c}{1+2c} - \frac{b c (2+c)}{(1+c)(1-b+c)}
\end{equation}
with $b=\sqrt{9\phi/8}$ and $c=11\phi/16$, and $\phi_0 = (4/3)^3/(7\ln 3 - 8 \ln 2 + 2)\approx 0.5718$. This theoretical expectation is in reasonable agreement with experimental data \cite{tokuyama:pre:1994} as well as Stokesian dynamics simulations available for a limited set of larger volume fractions \cite{foss:jfm:1999} [Fig.~\ref{fig:diffusion-mono}(b)].

To compare the MPCD simulations with Eq.~\eqref{eq:diffusion-mono}, I performed 100 additional simulations of a single particle to measure $D_{0,1}$ (Fig.~S10 and Table \ref{tab:single-particle}). The simulated values of $D_{0,1}$ were consistently smaller than the theoretical expectation for a sphere with no-slip boundary conditions, $D_{0,1} = k_{\rm B}T/(3 \pi \eta_0 d_1)$, with errors of 15.3\%, 0.7\%, and 4.0\% for diameters $3\,\ell$, $6\,\ell$, and $12\,\ell$, respectively. The simulated suspension long-time self-diffusion coefficients $D_1$ were somewhat smaller than expected from Eq.~\eqref{eq:diffusion-mono} at small volume fractions but then tended to be somewhat larger at large volume fractions [Fig.~\ref{fig:diffusion-mono}(a)]. This weaker volume-fraction dependence is clearer when the simulated $D_1$ is normalized by the simulated $D_{0,1}$ [Fig.~\ref{fig:diffusion-mono}(b)]: all three diameters gave $D_1/D_{0,1}$ similar to each other, and all had a somewhat weaker dependence on $\phi$ than expected from Eq.~\eqref{eq:diffusion-mono}.
\begin{table}
  \centering
  \caption{Long-time self-diffusion coefficient $D_{0,1}$ and sedimentation velocity $U_{0,1}$ for a single particle with different diameter $d_1$. The applied force $f_1$ in the sedimentation simulations was proportional to the particle volume with $f_1 = 0.5\,\varepsilon/\ell$ for $d_1 = 6\,\ell$. The simulated values are compared to theoretical expectations for a sphere with no-slip boundary conditions.}
  \begin{tabular}{ccccccc}
  $d_1$ ($\ell$) & \multicolumn{2}{c}{$D_{0,1}$ ($10^{-3}\ell^2/\tau$)} & \multicolumn{2}{c}{$U_{0,1}$ ($10^{-3}\ell/\tau$)} \\
  & expected & simulated & expected & simulated \\ \hline
  3 & 8.94 & $7.57 \pm 0.23$ & 0.56 & $0.48 \pm 0.03$\\  
  6 & 4.47 & $4.44 \pm 0.16$ & 2.24 & $2.09 \pm 0.03$\\
  12 & 2.24 & $2.15 \pm 0.07$ & 8.94 & $8.67 \pm 0.03$
  \end{tabular}
  \label{tab:single-particle}
\end{table}

\begin{figure}
    \centering
    \includegraphics{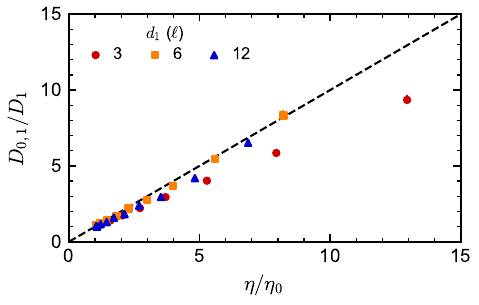}
    \caption{Test of the generalized Stokes--Einstein relationship between the simulated shear viscosity $\eta$ and long-time self-diffusion coefficient $D_1$ for monodisperse suspensions of particles with diameter $d_1$ using the data from Figs.~\ref{fig:viscosity-mono} and \ref{fig:diffusion-mono}. The dashed line, $D_{0,1}/D_1 = \eta/\eta_0$, is the theoretical expectation.}
    \label{fig:stokes-einstein}
\end{figure}

In previous work \cite{wani:jcp:2022, wani:sm:2024}, I did not measure the suspension shear viscosity so I assumed that the shear viscosity and long-time self-diffusion coefficient satisfied the generalized Stokes--Einstein relationship \cite{mason:prl:1995}, $\eta/\eta_0 = D_{0,1}/D_1$, to apply finite-size corrections. Since $D_1$ and $\eta$ have both been measured here, I tested the validity of this assumption by comparing the proportionality of the two quantities (Fig.~\ref{fig:stokes-einstein}). I found that the generalized Stokes--Einstein relationship is well-satisfied for diameters $6\,\ell$ and $12\,\ell$, but there was some discrepancy for diameter $3\,\ell$. The previous work used particles with diameter $6\,\ell$, so the Stokes--Einstein assumption seems to be justified in that case.

I last measured the sedimentation velocity $U_1$ of the monodisperse suspensions under the applied force described in Sec.~\ref{sec:methods:sediment}. The static structure factor at zero wavevector, $S(0)$, was needed to apply finite-size corrections to $U_1$, so I extrapolated it from the measured $S(q)$ (Fig.~S1). I confirmed that, as expected, the simulated values of $S(0)$ were the same for all three particle diameters (Fig.~S2). The values of $S(0)$ also matched the theoretical expectation based on the isothermal compressibility of the Carnahan--Starling equation of state \cite{carnahan:jcp:1969}. This agreement justifies the use of the Carnahan--Starling equation of state to also compute $g^+$ in Eq.~\eqref{eq:viscosity-mono}. The simulated values of $S(0)$ were used to apply the finite-size corrections to $U_1$.

\begin{figure}
    \centering
    \includegraphics{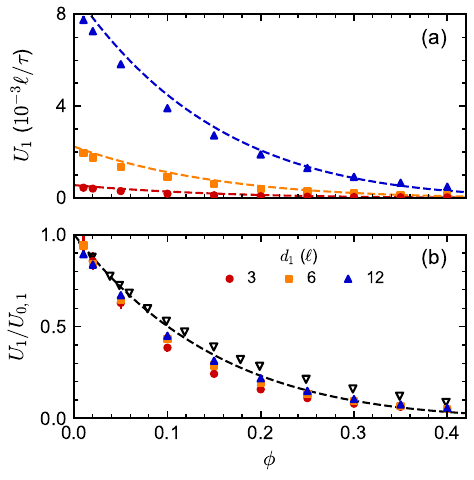}
    \caption{(a) Simulated sedimentation velocity $U_1$ as a function of volume fraction $\phi$ for monodisperse suspensions of particles with diameter $d_1$. The dashed line is Eq.~\eqref{eq:sediment-mono} with $U_{0,1}$ set to its theoretically expected value. The applied force $f_1$ was proportional to the particle volume with $f_1 = 0.5\,\varepsilon/\ell$ for $d_1 = 6\,\ell$. (b) The same as (a) but the simulated $U_1$ is normalized by the simulated single-particle sedimentation velocity $U_{0,1}$. The open symbols are the Stokesian dynamics results of Ref.~\citenum{wang:jcp:2015}.}
    \label{fig:sediment-mono}
\end{figure}

I then compared the sedimentation velocities with a semi-empirical approximation for $U_1$ \cite{wang:jcp:2015},
\begin{equation}
\frac{U_1}{U_{0,1}} = (1-\phi)^{6.5464},
\label{eq:sediment-mono}
\end{equation}
where $U_{0,1}$ is the sedimentation velocity of a single particle. Note that the first-order series expansion of Eq.~\eqref{eq:sediment-mono} matches Batchelor's well-known result for the dilute limit \cite{batchelor:jfm:1972} within rounding to two digits. It is also in good agreement with Stokesian dynamics simulations for modest volume fractions \cite{wang:jcp:2015}, with those simulations giving somewhat larger values of $U_1$ at larger volume fractions (Fig.~\ref{fig:sediment-mono}).

To compare the MPCD simulations with Eq.~\eqref{eq:sediment-mono}, I performed sedimentation simulations of a single particle (800 simulations for diameter $3\,\ell$, 100 simulations for diameters $6\,\ell$ and $12\,\ell$) to measure $U_{0,1}$ (Table \ref{tab:single-particle}). As for the self-diffusion coefficient, the simulated values of $U_{0,1}$ were typically smaller than theoretically expected for a sphere with no-slip boundary conditions, $U_{0,1} = f_1/(3 \pi \eta_0 d_1)$, with errors of 14.3\%, 6.7\%, 3.0\% for diameters $3\,\ell$, $6\,\ell$, and $12\,\ell$, respectively. The simulated $U_1$ for the suspensions were then also smaller than expected based on Eq.~\eqref{eq:sediment-mono} [Fig.~\ref{fig:sediment-mono}(a)] for most volume fractions, with some crossover at the largest volume fractions. Interestingly, after normalizing by the simulated $U_{0,1}$, the simulated $U_1$ had a somewhat stronger volume fraction dependence than expected [Fig.~\ref{fig:sediment-mono}(b)] at smaller volume fractions, although the self-diffusion coefficient had a weaker dependence. This dependence became weaker than expected for larger volume fractions.

Overall, particles with diameter $3\,\ell$ exhibited the largest deviations from theoretical expectations for the suspension viscosity $\eta$ as well as the long-time self-diffusion coefficient $D_{0,1}$ and sedimentation velocity $U_{0,1}$ of a single particle, but $D_1/D_{0,1}$ and $U_1/U_{0,1}$ still had a similar volume-fraction dependence as the other particles. The particles with diameter $3\,\ell$ also had the largest departure from the generalized Stokes--Einstein relationship between $D_1$ and $\eta$. These errors for diameter $3\,\ell$ might be related to known artifacts in MPCD when the ratio of the diameter of a colloidal particle to the edge length of the collision cell becomes small \cite{padding:pre:2006}, as the edge length of the collision cell is roughly the length scale over which hydrodynamics can be resolved \cite{huang:pre:2012}. Particles with diameters $6\,\ell$ and $12\,\ell$ gave results that were more closely aligned with theoretical expectations, but there were still some differences between the two particle sizes. These differences might then be related to small differences in their discrete representations. Together, these discrepancies highlight the potential sensitivity of the MPCD method to discretization effects, which should be carefully considered. However, the results for all particle sizes seemed to be reasonably consistent with each other, particularly after normalizing by values for a single particle, and with theoretical expectations so I proceeded to simulate bidisperse suspensions.

\subsection{Bidisperse suspensions}
\label{sec:results:bi}
\begin{figure}
    \centering
    \includegraphics{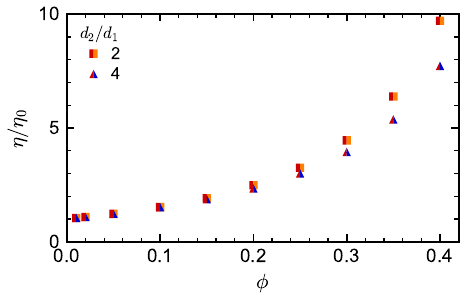}
    \caption{Simulated shear viscosity $\eta$ as a function of volume fraction $\phi$ for bidisperse suspensions of particles with diameters $d_1 = 3\,\ell$ and varied $d_2$. The suspensions had equal amounts of each particle by volume.}
    \label{fig:viscosity-bi}
\end{figure}

Following the same workflow as for the monodisperse suspensions, I first measured the viscosity $\eta$ of the bidisperse suspensions in the low-shear limit (Fig.~\ref{fig:viscosity-bi}). The viscosity had a similar dependence on volume fraction for both diameter ratio 2 and 4, with $\eta$ for diameter ratio 2 increasing somewhat faster with $\phi$. This behavior is qualitatively consistent with models for the shear viscosity of polydisperse suspensions based on effective volume fraction \cite{luckham:jcis:1999, qin:jcis:2003, shewan:jnfm:2015} because the maximum attainable suspension volume fraction, at which the shear viscosity diverges in these models, increases with diameter ratio \cite{anzivino:jcp:2023}. However, I note that the simulated shear viscosity was larger than predicted by these models for the larger volume fractions. Further, $\eta$ for the monodisperse suspension with diameter $6\,\ell$ was larger than for diameter $12\,\ell$, although both should be the same. I attributed these differences to discretization artifacts in Sec.~\ref{sec:results:mono}, and they may contribute here to differences in the $\phi$-dependence of the bidisperse suspension shear viscosity.

\begin{figure*}
    \centering
    \includegraphics{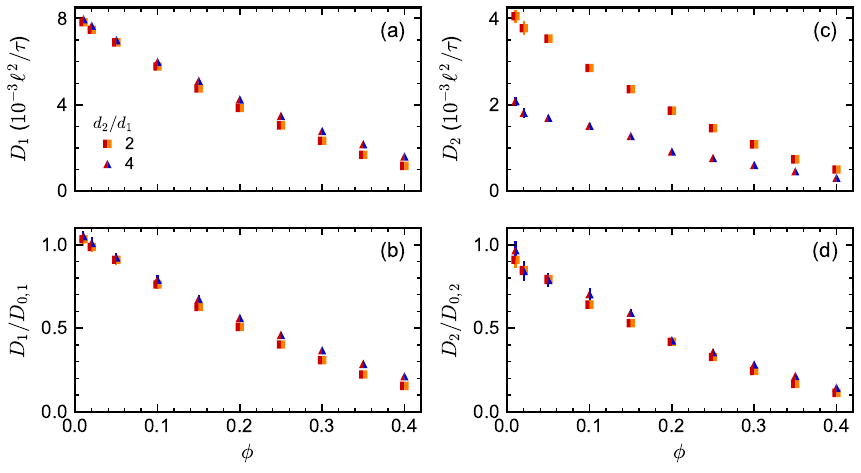}
    \caption{Simulated long-time self-diffusion coefficients (a) $D_1$ and (c) $D_2$ as a function of volume fraction $\phi$ for bidisperse suspensions of particles with diameter $d_1 = 3\,\ell$ and varied $d_2$. The suspensions had equal amounts of each particle by volume. (b) and (d) are the same as (a) and (c) but $D_1$ and $D_2$ are normalized by the simulated single-particle self-diffusion coefficients $D_{0,1}$ and $D_{0,2}$.}
    \label{fig:diffusion-bi}
\end{figure*}

I next characterized the long-time self-diffusion coefficients $D_1$ and $D_2$ for both particle sizes in the bidisperse suspensions. Regardless of the size of the larger particle $d_2$, $D_1$ had nearly the same value for the particles with $d_1 = 3\,\ell$ [Fig.~\ref{fig:diffusion-bi}(a)--(b)]. As expected, $D_2$ was about twice as large for $d_2 = 6\,\ell$ than for $d_2 = 12\,\ell$ [Fig.~\ref{fig:diffusion-bi}(c)]; however, $D_2/D_{0,2}$ was nearly the same for both diameter ratios [Fig.~\ref{fig:diffusion-bi}(d)]. This collapse is qualitatively consistent with the particles following a generalized Stokes--Einstein relationship with a similar suspension viscosity. I plotted $D_{0,i}/D_i$ against $\eta$ for the bidisperse suspensions (Fig.~S9), and I confirmed that the two were roughly linearly proportional for both types of particles. However, as for the monodisperse suspensions, the larger particles had the proportionality expected from the generalized Stokes--Einstein relationship, but the smaller particles had a smaller proportionality constant than expected.

\begin{figure*}
    \centering
    \includegraphics{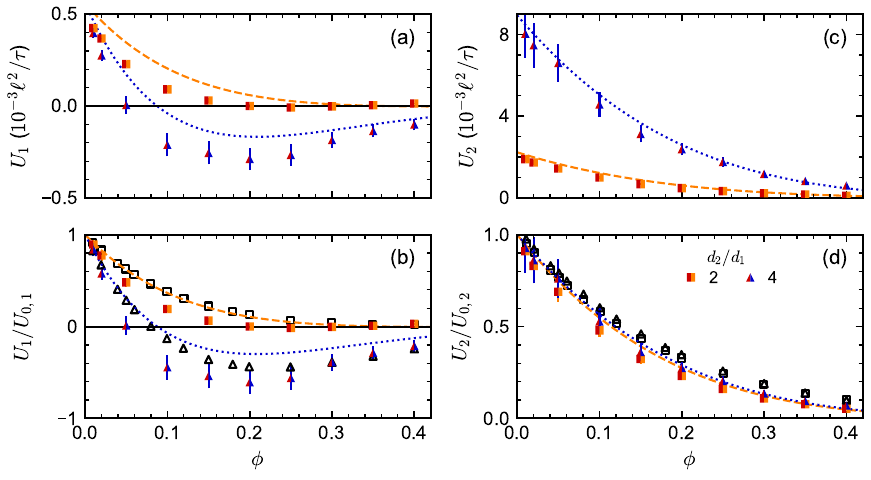}
    \caption{Simulated sedimentation velocities (a) $U_1$ and (c) $U_2$ as a function of volume fraction $\phi$ for bidisperse suspensions of particles with diameter $d_1 = 3\,\ell$ and varied $d_2$. The suspensions had equal amounts of each particle by volume, and the applied force $f_1$ was proportional to the particle volume with $f_1 = 0.5\,\varepsilon/\ell$ for $d_1 = 6\,\ell$. (b) and (d) are the same as (a) and (c) but $U_1$ and $U_2$ are normalized by the simulated single-particle sedimentation velocities $U_{0,1}$ and $U_{0,2}$. The dashed and dotted lines are Eqs.~\eqref{eq:sediment-bi:u1} and \eqref{eq:sediment-bi:u2} for $d_2/d_1 = 2$ and 4, respectively, using the theoretically expected values of $U_{0,1}$ and $U_{0,2}$. The open symbols are the Stokesian dynamics results of Ref.~\citenum{wang:jcp:2015}.}
    \label{fig:sediment-bi}
\end{figure*}

I last considered the sedimentation velocities $U_1$ and $U_2$ for both particle sizes in the bidisperse suspensions. I extrapolated the partial static structure factors at zero wavevector $S_{ij}(0)$ from the simulated $S_{ij}(q)$ (Figs.~S5 and S6) and used these values to apply finite-size corrections. An approximation of the sedimentation velocity in polydisperse suspensions has been developed \cite{davis:aichej:1994},
\begin{align}
\frac{U_1}{U_{0,1}} &= (1 - \phi)^{c_{11}} [1 + (c_{11} - c_{12}) \phi/2] \label{eq:sediment-bi:u1} \\
\frac{U_2}{U_{0,2}} &= (1 - \phi)^{c_{22}} [1 + (c_{22} - c_{21}) \phi/2]
\label{eq:sediment-bi:u2}
\end{align}
with $c_{11} = c_{22} = 6.5464$, while $c_{12} = \{11.966, 29.392\}$ and $c_{21} = \{4.7167, 4.0146\}$ for $d_2/d_1 = \{2,4\}$ \cite{wang:jcp:2015}. This approximation is in reasonable agreement with Stokesian dynamics simulation results \cite{wang:jcp:2015} [Fig~\ref{fig:sediment-bi}(b) and (d)].

As for the monodisperse suspensions, the sedimentation velocities $U_1$ and $U_2$ simulated using MPCD were consistently somewhat smaller than theoretically expected [Fig.~\ref{fig:diffusion-bi}(a) and (c)]. The sedimentation velocity for the larger particles with diameters $6\,\ell$ or $12\,\ell$ was typically in somewhat better agreement than for the smaller particles with diameter $3\,\ell$; this result is sensible given the issues noted for the smallest particles in the monodisperse suspensions. However, good agreement with both theoretical expectations and Stokesian dynamics simulations was found after normalization by the simulated values for a single particle [Figs.~\ref{fig:sediment-bi}(b) and (d)].

It is particularly noteworthy that the MPCD simulations correctly capture the reversal in direction of $U_1$ for diameter ratio 4, even achieving a fastest speed at a similar volume fraction as expected from Eq.~\eqref{eq:sediment-bi:u1}. Qualitatively, this reversal is driven by solvent backflow induced by the sedimentation of the larger particles, so the MPCD simulations are reasonably capturing this hydrodynamic behavior. There are some quantitative differences that, based on the monodisperse suspensions, might be resolved by using larger particles through proportional increases in both $d_1$ and $d_2$. Unfortunately, I was not able to test this at present due to computational cost.

\section{Conclusions}
\label{sec:conclusions}
Using MPCD with a discrete particle model, I have simulated the shear viscosity, long-time self-diffusion coefficient, and sedimentation velocity in monodisperse and bidisperse hard-sphere colloidal suspensions as a function of volume fraction for three different sizes of particles. The results for the monodisperse suspensions were compared to theoretical expectations and prior simulations, generally finding reasonable agreement. Discrepancies between results for the different particle sizes highlight the potential for discretization artifacts in MPCD from both the size of the particle relative to the collision cell and the representation of the particle surface. The sedimentation velocities for the bidisperse suspensions were also in reasonable agreement with theoretical expectations and prior simulations.

These simulations provide useful reference data on the transport properties of  colloidal suspensions that supplements existing literature. This work also establishes the suitability of the MPCD method for simulating suspensions containing multiple types of particles. MPCD is a convenient method for simulating hydrodynamic interactions near solid boundaries as well as a variety of soft materials, not only particles, because of its simple particle-based solvent. This work lays the groundwork for using MPCD to simulate more complex dynamic processes for colloidal suspensions. For example, MPCD might be used to simulate the drying-induced assembly of size-stratified particle coatings, as well as for measuring transport coefficients that can be used to model this process at a continuum level.

\section*{Supplementary Material}
See the supplementary material for measured and extrapolated static structure factors, velocities in reverse nonequilibrium simulations, and time derivatives of mean squared displacements for all suspensions at representative volume fractions.

\section*{Conflicts of interest}
The author has no conflicts to disclose.

\section*{Data Availability}
The data that support the findings of this study are available from the author upon reasonable request.

\section*{Acknowledgments}
This material is based upon work supported by the National Science Foundation under Award No.~2310724 and the International Fine Particle Research Institute. This work used Delta at the National Center for Supercomputing Applications through allocation CHM250006 from the Advanced Cyberinfrastructure Coordination Ecosystem: Services \& Support (ACCESS) program \cite{access}, which is supported by National Science Foundation grants \#2138259, \#2138286, \#2138307, \#2137603, and \#2138296. This work was also completed with resources provided by the Auburn University Easley Cluster.

\bibliography{references}

\end{document}


\title{Supplementary material for ``Transport properties of monodisperse and bidisperse hard-sphere colloidal suspensions from multiparticle collision dynamics simulations''}

\author{Michael P. Howard}
\email{mphoward@auburn.edu}
\affiliation{Department of Chemical Engineering, Auburn University, Auburn, AL 36849, USA}

\maketitle

\section{Monodisperse suspensions}
\begin{figure}[!h]
    \centering
    \includegraphics{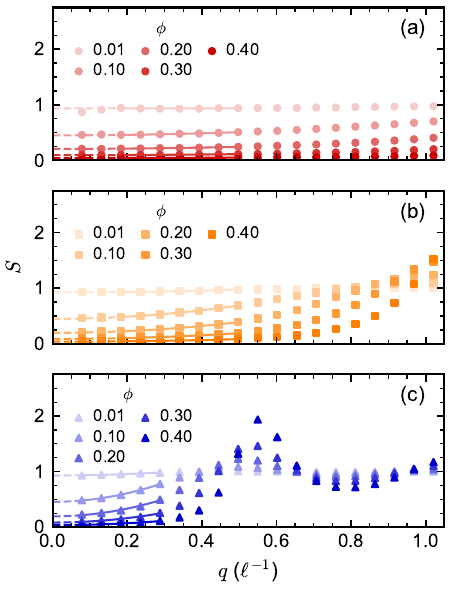}
    \caption{Static structure factor $S$ as a function of wavevector $q$ for monodisperse suspensions of particles with diameter (a) $3\,\ell$, (b) $6\,\ell$, and (c) $12\,\ell$ at selected volume fractions $\phi$. The solid lines show the fit described in the main text, while the dashed lines show the extrapolation to $q=0$.}
    \label{fig:sk-mono}
\end{figure}

\begin{figure}[!h]
    \centering
    \includegraphics{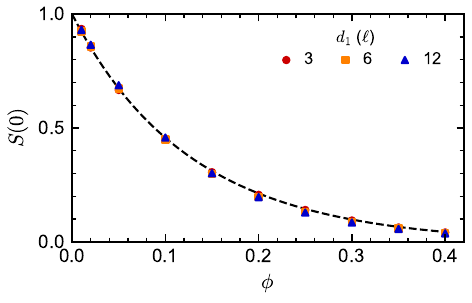}
    \caption{Static structure factor extrapolated to $q = 0$, $S(0)$, for monodisperse suspensions of particles with diameter $d_1$. The dashed line is the theoretical expectation for the Carnahan--Starling equation of state, $S(0) = (1-\phi)^4 / (1+4\phi+4\phi^2-4\phi^3+\phi^4)$.}
    \label{fig:s0-mono}
\end{figure}

\begin{figure}[!h]
    \centering
    \includegraphics{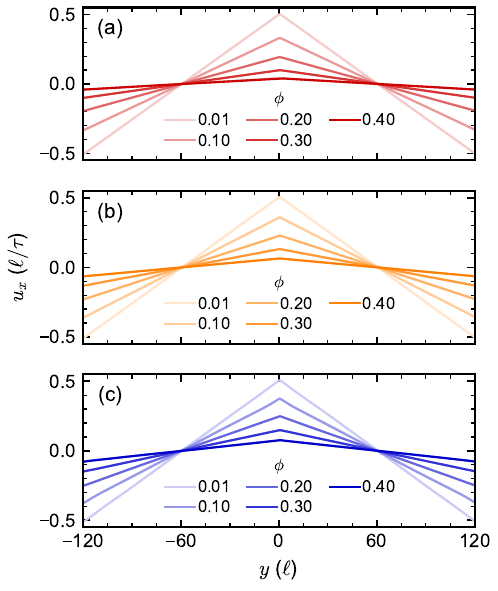}
    \caption{Mass-averaged velocity $u_x$ as a function of position $y$ for monodisperse suspensions of particles with diameter (a) $3\,\ell$, (b) $6\,\ell$, and (c) $12\,\ell$ at selected volume fractions $\phi$ in reverse nonequilibrium simulations.}
    \label{fig:velocity-mono}
\end{figure}

\begin{figure}[!h]
    \centering
    \includegraphics{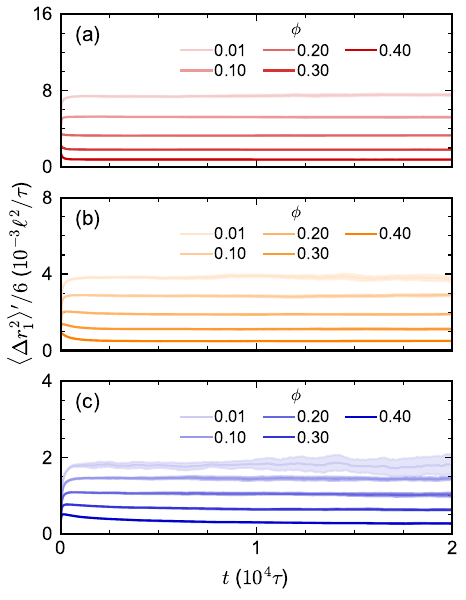}
    \caption{Time derivative $\langle \Delta r_1^2\rangle'$ of mean squared displacement $\Delta r_1^2$ as a function of time $t$ for monodisperse suspensions of particles with diameter (a) $3\,\ell$, (b) $6\,\ell$, and (c) $12\,\ell$ at selected volume fractions $\phi$.}
    \label{fig:msd-mono}
\end{figure}

\clearpage
\section{Bidisperse suspensions}
\begin{figure}[!h]
    \centering
    \includegraphics{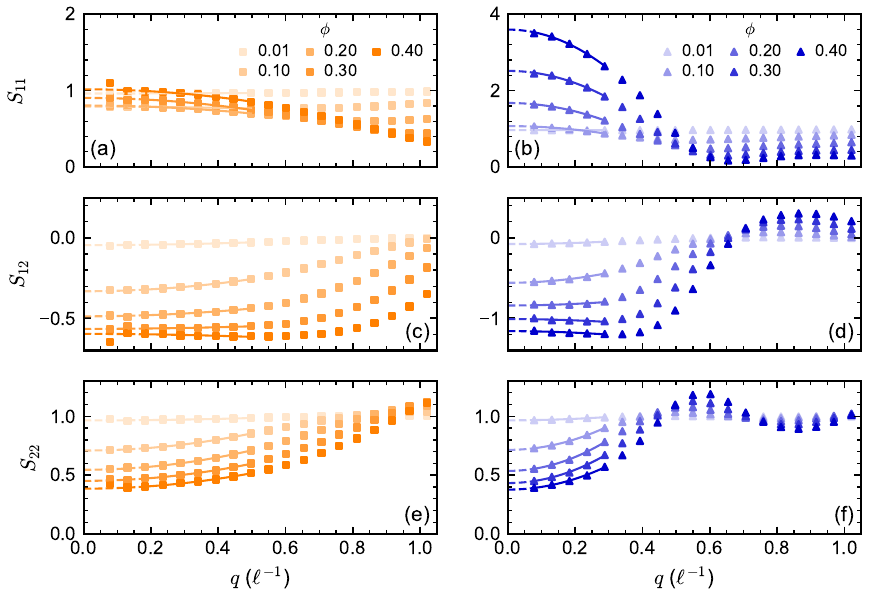}
    \caption{Partial static structure factors (a--b) $S_{11}$, (c--d) $S_{12}$, and (e--f) $S_{22}$ as a function of wavevector $q$ for bidisperse suspensions of particles with diameter $d_1 = 3\,\ell$ and particles with diameter (a, c, e) $d_2 = 6\,\ell$ or (b, d, f) $d_2 = 12\,\ell$ at selected volume fractions $\phi$. There are equal volumes of both particle types. The solid lines show the fit described in the main text, while the dashed lines show the extrapolation to $q=0$.}
    \label{fig:sk-bi}
\end{figure}

\begin{figure}[!h]
    \centering
    \includegraphics{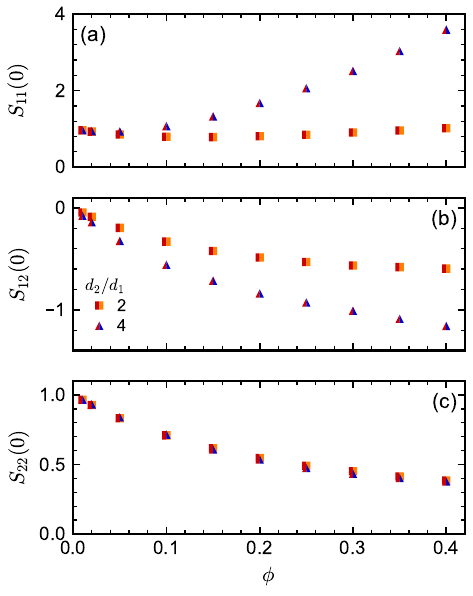}
    \caption{Partial static structure factors extrapolated to $q = 0$, (a) $S_{11}(0)$, (b) $S_{12}(0)$, and (c) $S_{22}(0)$, for bidisperse suspensions of particles with diameters $d_1 = 3\,\ell$ and varied $d_2$ at selected volume fractions $\phi$. There are equal volumes of both particle types.}
    \label{fig:s0-bi}
\end{figure}

\begin{figure}[!h]
    \centering
    \includegraphics{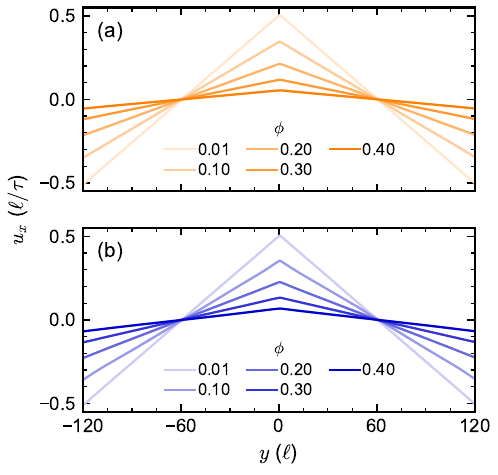}
    \caption{Mass-averaged velocity $u_x$ as a function of position $y$ for bidisperse suspensions of particles with diameter $d_1 = 3\,\ell$ and particles with diameter (a) $d_2 = 6\,\ell$ or (b) $d_2 = 12\,\ell$ at selected volume fractions $\phi$ in reverse nonequilibrium simulations. There are equal volumes of both particle types.}
    \label{fig:velocity-bi}
\end{figure}

\begin{figure}[!h]
    \centering
    \includegraphics{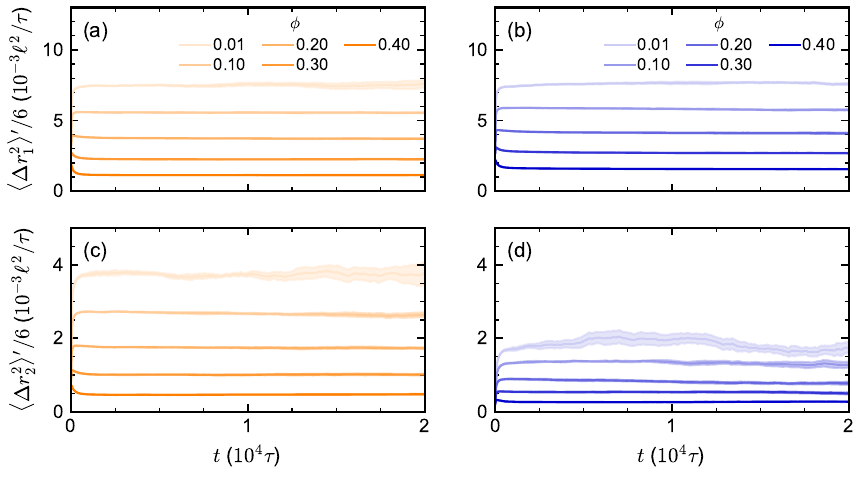}
    \caption{Time derivative $\langle \Delta r_i^2\rangle'$ of mean squared displacement $\Delta r_i^2$ for particles of type $i$ as a function of time $t$ for bidisperse suspensions of particles with diameter $d_1 = 3\,\ell$ and particles with diameter (a, c) $d_2 = 6\,\ell$ or (b, d) $d_2 = 12\,\ell$ at selected volume fractions $\phi$. There are equal volumes of both particle types.}
    \label{fig:msd-bi}
\end{figure}

\begin{figure}
    \centering
    \includegraphics{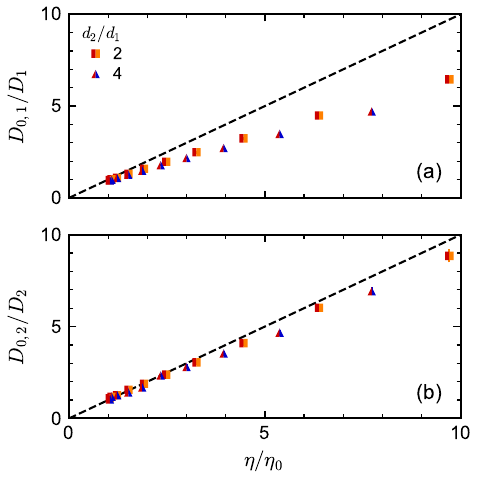}
    \caption{Test of the generalized Stokes--Einstein relationship between the simulated shear viscosity $\eta$ and long-time self-diffusion coefficients (a) $D_1$ and (b) $D_2$ for bidisperse suspensions of particles with diameter $d_1 = 3\,\ell$ and varied $d_2$ using the data from Figs.~6 and 7. The dashed line, $D_{0,i}/D_i = \eta/\eta_0$, is the theoretical expectation.}
    \label{fig:stokes-einstein}
\end{figure}

\clearpage
\section{Single particles}
\begin{figure}[!h]
    \centering
    \includegraphics{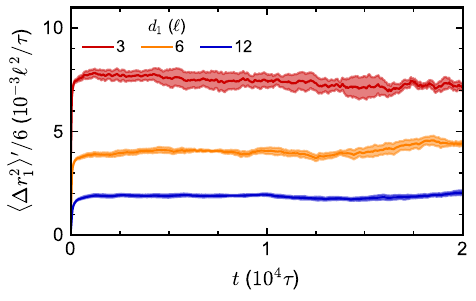}
    \caption{Time derivative $\langle \Delta r_1^2\rangle'$ of mean squared displacement $\Delta r_1^2$ as a function of time $t$ for a single particle with diameter $d_1$.}
    \label{fig:msd-single}
\end{figure}